# On The Relativity of Redshifts: Does Space Really "Expand"?

**Geraint F. Lewis**

**Sydney Institute for Astronomy, School of Physics, A28**
**The University of Sydney, NSW 2006**

*In classes on cosmology, students are often told that photons stretch as space expands, but just how physical is this picture? Does space really expand? In this article, we explore the notion of the redshift of light within Einstein's general theory of relativity, showing that the core underpinning principles reveal that redshifts are both simpler and more complex than you might naively think. This has significant implications for the observed redshifting of photons as they travel across the universe, often referred to as the cosmological redshift, and for the idea of expanding space.*

## Stretching Photons

In an expanding universe, the light from distant galaxies is redshifted, with the wavelength of observed spectral features being longer than those measured in the laboratory. To anyone who has taken an undergraduate course on cosmology, the source of this redshifting is obvious, having been told that photons "stretch" as the space expands. This statement is often accompanied with a picture like Figure 1, with a blue photon stretched into a red photon as space expands during its journey between two cosmological observers.

All of this is pretty satisfying, and life can happily continue. But with a little more thought, a few niggling issues appear. If expanding space can stretch a photon, a photon that is extremely tiny, is expanding space stretching atoms and molecules? Is expanding space stretching stars and galaxies? And are Brooklyn and its inhabitants expanding with the universe, as discussed in the wonderful scene in Woody Allen's "Annie Hall". When faced with such questions, you may turn to Google and find out what the experts have to say, and you may find yourself rather surprised.

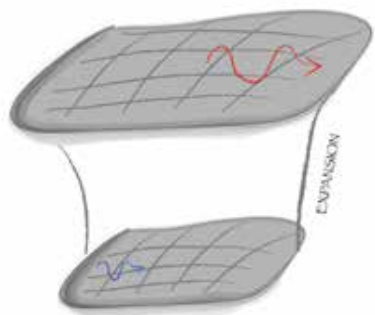

**Figure 1:** Typical diagram demonstrating how expanding space stretches photons as they travel across the universe.

John Peacock, author of "Cosmological Physics", attacks the misconceptions in cosmology, noting that "[t]he worst of these is the 'expanding space' fallacy" [1]. But Peacock is just one cosmologist, and you may turn to others for further scientific insight, but you'll find no solace there. Cosmological giants, Martin Rees and Steven Weinberg, tell us

> "…how is it possible for space, which is utterly empty, to expand? How can nothing expand? The answer is: space does not expand. Cosmologists sometimes talk about expanding space, but they should know better."

So experts tell us that space doesn't expand! Just what is the layperson to make of this? And if space doesn't expand, just what stretches a photon traveling across the universe? To start to answer these questions, we need to take a step back and really understand the mechanism of the redshifting of light in a relativistic universe.

## Three Types of Redshift?

When flipping through a physics textbook, students are typically told that there are three different redshifts seen within Einstein's relativity, each applicable in particular circumstances. These are;

**Doppler Redshift:** first encountered in the flat space-time of special relativity, this concerns the observation of photons by observers who are moving relative to one another.

**Gravitational Redshift:** a classical consequence of general relativity, observers at different locations in a gravitational field measure different wavelengths when exchanging photons.

**Cosmological Redshift:** a staple of cosmology classes, this is the case where observers exchange photons over cosmological distances in an expanding universe.

These appear to be distinct physical processes, and governed by quite different equations. But let's again ask ourselves the mechanism by which the redshifting occurs. We've already seen what students are told that in the cosmological case. In the case of the gravitational redshift, photons apparently lose energy as they climb out of a gravitational potential.

But what about the first case considered above, the Doppler shifting of special relativity? Just where does the redshifting occur in this scenario? Understanding this is key to understanding relativistic redshifts in general. But let's start with a photon moving in a gravitational field

## Of Gravity and Rockets

As already mentioned, the gravitational redshift appears to occur as photons lose energy as they climb in a gravitational field, a situation we can represent schematically shown in Figure 2.

Considered one of the classical test of general relativity, this phenomenon was experimentally verified in 1959 by Robert Pound and Glen Rebka in the Harvard tower experiment, where photons were sent on journeys up and down a 22m path and their energies measured, finding precise agreement with the predictions of general relativity. This was a particularly difficult experiment, mainly due to the weakness of the Earth's gravitational field, but using the Mossbauer effect, where the emitting and absorbing atoms are locked into a crystal lattice, allowed the extremely fine measurement of the photon energies and hence the redshift. But let's not worry about the messiness of experimental physics and instead consider the theoretical aspects of gravitational redshifting.

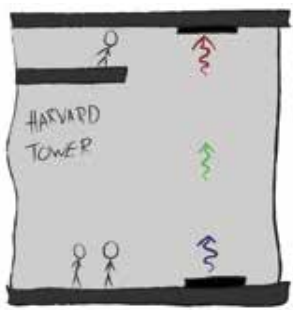

**Figure 2:** Schematic representation of the famous Harvard Tower experiment of Pound and Rebka, showing photons are redshifted when then travel in a gravitational field.

Let's start with an initially blue photon on an upward journey in a gravitational field. For a significantly large change in the gravitational potential, the detected photon at the end of the journey will be red. But where does the redshifting occur? It seems to be that this is a continuous effect on the photon as it travels, with each step upwards robbing the photon of a little bit more energy. Hence, in the representation above, the intermediate photon, the one half way along in its journey, is green. This seems to make intuitive sense, but the story does not end here.

Let's take a further step back to one of the founding principles of general relativity; in particular what Einstein called the "happiest thought in my life". This was the realisation that for someone in free fall, the gravitational field vanishes; as a trip in the "vomit-comet" demonstrates, all those in free-fall float around like astronauts in deep space. More formally, this is known as the "Equivalence Principle" and can be stated that no physical experiment can reveal to an observer (with no visual clues) whether they are floating in deep space, far from sources of gravity, or in free fall in a gravitational fields, and this property, known as "local flatness", is one of the key features of the space-time of general relativity.

However, there is another side of the equivalence principle that will be useful here, namely that there is no physical experiment our observer, who still has no external visual clues, could do to distinguish between being at rest in a gravitational field or being inside a uniformly accelerating rocket in deep space. Throw a ball on the surface of the Earth, and throw an identical ball on a deep space rocket accelerating at 1-g, the resultant paths will be the same.

So, according to the equivalence principle, if we repeat the Harvard tower experiment in a rocket accelerating at 1-g, we should get an identical result, namely that a photon fired from the back of the ship should be at a lower energy when detected at the front of the ship (see Figure 3). In the following, we will consider an extreme acceleration (probably not conducive to comfortable spaceflight) such that the photon fired the back of the ship is blue, while that detected at the front of the ship is red. So what colour is the photon half way up the rocket?

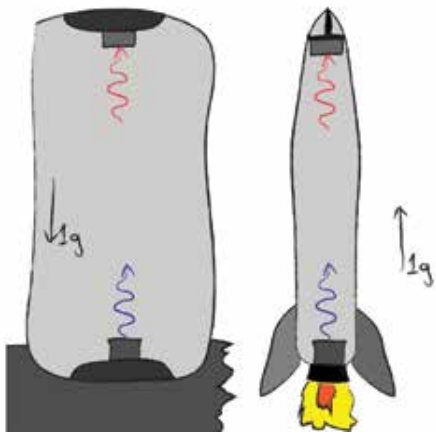

**Figure 3: According to the equivalence principle, repeating the Harvard Tower experiment in an accelerating rocket is deep space should yield the same results as on Earth.**

Let's turn our attention to some observers who are not accelerating, observers who are simply sitting in space, at rest with respect to each other. We can initially place the rocket at rest with these observers, with the engines ready to fire. The button is pressed and the rocket roars, and at the same instant the photon is fired from the base of the rocket. At this instant, as observers inside and outside of the rocket are at rest with each other, both measure this newly emitted photon as being blue.

Now, let's think of the photon halfway through its journey, traveling through the middle of the rocket. We know from the equivalence principle, that the situation on-board the rocket must be identical to those in a gravitational field and so an on-board observer would see this photon as being green. However, what does an external, at rest observer see? To these observers, the photon has simply travelled through empty, flat space-time, and an observer measuring the photon at the midpoint would find it unredshifted and as blue as when it was emitted. So is the photon blue or green?

How are we to reconcile this situation? Does this mean that the Equivalence Principle, one of the founding ideas of general relativity, breaks down? The answer is no, and the reason is that it matters who is observing the photon at the midpoint of the rocket.

Let's look at what happens once the photon has been emitted. The rocket accelerates as the photon travels, so compared to the observers at rest outside the ship, those inside are moving at high velocity when the pho-ton is traveling though the midpoint of the rocket. So, it should come as no surprise that they measure the energy of the photon to have a different value. And when the photon is absorbed at the top of the rocket, the relative velocity is even larger, and so while the external observers see the photon as still being blue, in a laboratory at the top of the rocket, the photon is now red (see Figure 4).

So, the key feature here is that the observed energy of a photon is a local thing, determined locally in an observer's laboratory, and the energy this depends upon what the photon and laboratory are doing. And given that we can analyse the situation in two apparently different ways, is there a more fundamental way of defining redshifts in relativity. The answer is yes!

But before we get to that, a little homework. Let's flip the situation and consider not a rocket in deep space, but observers in a gravitational field, some at rest, and some in free-fall. They repeat the Harvard tower experiment, and so those at rest see the photon redshifted as it climbs. What do the freefalling observers see? No calculations should be necessary!

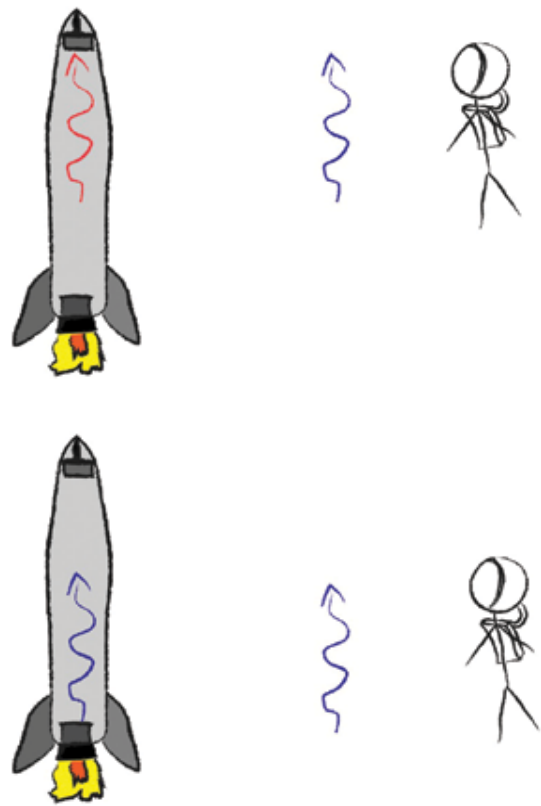

**Figure 4: What is the wavelength of a photon as seen by accelerating and non-accelerating observers?**

### Universal Redshifting

In 1994, Jayant Narlikar published a nice little paper in the American Journal of Physics titled "Spectral shifts in general relativity" [2], generalising some earlier work of John Synge in the early 1960s [3]. The central thrust of this paper is that it is incorrect to think that there are three distinct mechanisms for redshifting photons in relativity, and that there is truly only a single underlying mathematical description for use in all occasions.

Narlikar's paper is rather mathematical, but the basic idea is straightforward. In relativity, a photon is represented as a vector, a vector pointing in four-dimensional space-time. Unlike the nice vectors we are generally used to in classical physics, the magnitude of photon vectors is always zero, but they are mathematically very well behaved. Similarly, an observer's laboratory is defined by a collection of four-vectors (for those in the know, this is an orthonormal tetrad, or, if you want to sound very smart, a vierbein), each consisting of three pointing along the observer's spatial directions and one in their time direction. And to calculate the energy of a photon as seen by an observer in their laboratory, what we need to do is project the photon's four-vector on to the time component of the observer's coordinates (more technically, we take the vector dot-product between the two).

The dot-product of two vectors is done at a particular location so, as we expect, a photon measurement is a local thing, and we expect two laboratories with two different sets of laboratory four-vectors, will measure the same photon at the same location to have different energies. This local nature of the measurement of photon energies implies that the redshifting is something related to the properties of the observers, and the photon is not redshifted on its journey.

You may not like the above statement, as we know that in the curved space-time of general relativity, we have to "parallel-transport" our photon four-vector between our two observers; surely this is changing the photon as it travels? Let's go back to our rocket example. For our external observers, we can cover the space-time with the Minkowski metric of special relativity, allowing us to define the components of the photon's four-vector. But with this, these vector components do not change as the photon travels, and take the vector dot-product of this photon with observers in this space-time, be they stationary, moving with uniform velocity, or accelerating, reveals each sees a differing photon energy.

But, through the equivalence principle, we can explain the same scenario as being in a uniform gravitational field, and so can employ an appropriate space-time metric do describe this. In this metric, those originally on-board the rocket are at rest at different heights in the gravitational field, whereas our previously stationary observers are now in free-fall. As the photon travels in this coordinate system, the parallel propagation modifies the values of its four-vector, so these will be different at different location. But, again, taking the vector dot-products with the photon with observers reveals the same photon energies as before.

Remember, this is, physically, describing the same situation, and in one coordinate system the photon four-vector changes during its journey, whereas in the other the components do not. Does asking where does the redshifting of the photon occurs even mean anything? And what does the wavelength of a photon mean when there is no observer there to observe it?

### Does Space Really Expand?

After our journey around relativistic redshifts, we arrive back at the question we opened this article with, namely "Does space really expand?" As we have seen, the wavelength of a photon is not a unique thing, with the components of the photon four-vector dependent upon the choice of the metric to describe the underlying space-time, while the observed energy of a photon is dependent upon precisely what a particular observer is doing at the time they make the measurement. So, you should not think of the photon as travelling along with a little tag attached that records its wavelength. Wavelength is not a property of the photon, but of the "photon+observer" system.

So, let's look at the cosmological case in a little more detail. In a typical cosmology course, students are introduced to the Friedmann-Robertson-Walker metric to describe the space-time of an expanding homogeneous and isotropic universe, although they are not often told that this is not the only mathematical description of this space-time. But let's stick with the Friedmann-Robertson-Walker space-time for now. With this we typically consider a special group of observers, those at rest with regards to the coordinates of this metric, the so-called "co-moving observers".

Let's consider two of these special observers, A and B, separated by a large distance in a Friedmann-Robertson-Walker universe, and let's assume A sends a photon to B. We know that this photons wavelength will be stretched by the amount the universe has expanded during its journey. But let's also consider a myriad of additional observers, each at rest with regards to the Friedmann-Robertson-Walker coordinates, spread evenly between A and B, and as the photon passes, each will measure its energy. Each will see the wavelength of the photon as being progressively larger, with the photon apparently stretching in its journey.

However, if we consider one of the intermediate observers, we can ask what they see. To them, their adjacent observers are moving away in locally flat space-time, and that the redshifting they see is simply the Doppler shift due to motion. So the entire redshift between A and B can be considered just a long series of Doppler shifts. But, again, this is difficult to visualise without inserting our long chain of observers into the picture.

Hence, we arrive at the crux of this article, namely that the concept of expanding space is useful in a particular scenario, considering a particular set of observers, those "co-moving" with the coordinates in a space-time described by the Friedmann-Robertson-Walker metric, where the observed wavelengths of photons grow with the expansion of the universe. But we should not conclude that space must be really expanding *because* photons are being stretched. With a quick change of coordinates, expanding space can be extinguished, replaced with the simple Doppler shift .

While it may seem that railing against the concept of expanding space is somewhat petty, it is actually important to set the scene straight, especially for novices in cosmology. One of the important aspects in growing as a physicist is to develop an intuition, an intuition that can guide you on what to expect from the complex equation under your fingers. But if you assuming that expanding space is something physical, something like a river carrying distant observers along as the universe expands, the consequence of this when considering the motions of objects in the universe will lead to radically incorrect results.

So, what are the take home messages from this article? The first should be that the concept of redshifting in relativity is simpler than most textbooks portray, with a single underlying mathematical framework in which you can calculate the redshift in all cases. The second is that the concept of redshifting in relativity is more complex than most textbooks portray, as redshifting is not necessarily something that happens to a photon, but has more to do with what is happening to observers at the points of emission and absorption of a photon. But on the positive side, this should help students reinforce the concept that no particular metric, motion or location is unique or special when considering a situation. And the final message should address the concept of expanding space, that staple of cosmology textbooks. If all you want is an analogy to picture a photon traveling between two special observers, co-moving with the expansion in a Friedmann-Robertson-Walker metric, so that the wavelength expands with the universe, then you can talk about expanding space. But once you have finished, you should consider the contents of this article, and remind your students that while the picture seems comfortable and intuitive, it is no more than a picture and should be handled with care [4] [5].

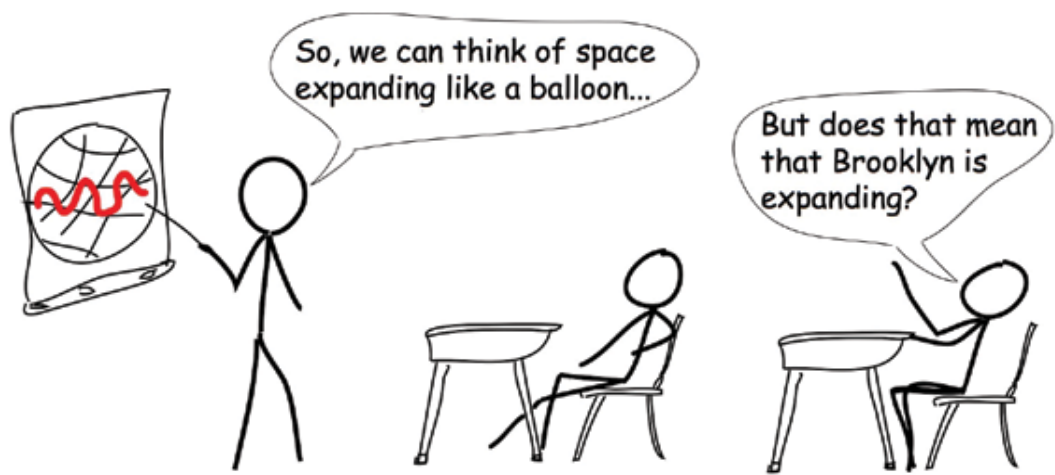

**Expanding space, a useful cosmological picture. But a picture none-the-less. Don't push it to too hard!**

## AUTHOR BIOGRAPHY

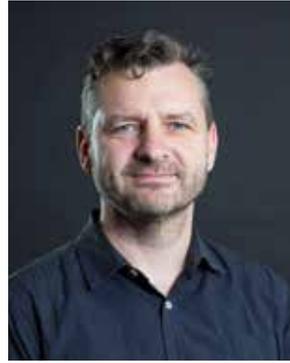

Geraint F. Lewis is a professor of astrophysics at the Sydney Institute for Astronomy, part of the University of Sydney's School of Physics. His research encompasses both theoretical and observational aspects of astrophysics, focusing upon cosmology, gravitational lensing and galactic cannibalism. Originally from Old South Wales, he arrived in Australia in 2000, joining the University of Sydney in 2002. He has published more than 250 papers on his research, and his book on the fine-tuning of the laws of physics for life, co-authored with Luke Barnes, will be published by Cambridge University Press in late 2016.

## Conferences 2016

**3–7 July 2016**
10th International Conference on Residual Stresses (ICRS-10)
Novotel Brighton Beach, Sydney
www.ansto.gov.au/Events/ICRS2016/index.htm

**4–5 July 2016**
Energy Future Conference and Exhibitions 2016. UNSW, Sydney.
www.ozenergyfuture.com

**4–8 July 2016**
International Conference on Supersymmetry and Unification of Fundamental Interactions (SUSY 2016). University of Melbourne.
indico.cern.ch/event/443176/

**10-15 July 2016**
15th international workshop on the physics of compressible turbulent mixing. University of Sydney
sydney.edu.au/engineering/events/iwpctm15

**10–12 July 2016**
Astrobiology Australasia 2016. CSIRO Perth
http://www.aa-meeting2016.com/

**11–15 July 2016**
NUSOD—2016 International Conference on Numerical Simulation of Optoelectronic Devices. University of Sydney
http://www.nusod.org/2016/

**18–22 July 2016**
The Multi-Messenger Astrophysics of the Galactic Centre
Palm Cove, Queensland
galacticcentre.space/

**20–24 July 2016**
LHP V: 5th International Workshop on Lattice Hadron Physics
Cairns Colonial Club Resort, Cairns, Queensland
www.physics.adelaide.edu.au/cssm/workshops/LHP5/

**5-8 September 2016**
OSA Congress: Photonics and Fiber Technology incorporating Australian Conference on Optical Fibre Technology (41st ACOFT), Bragg Gratings, Photosensitivity and Poling in Glass Waveguides (BGPP), and Nonlinear Photonics (NP).
SMC Conference & Function Centre, Sydney, Australia
http://www.osa.org/en-us/meetings/optics_and_photonics_congresses/photonics_and_fiber_technology/

**5-9 September 2016**
Topological matter, strings, and K-theory conference
The University of Adelaide
research.amsi.org.au/events/event/topological-matter-strings-and-k-theory/

**11–16 September 2016**
International Conference on Nuclear Physics (INPC2016)
Adelaide Convention Centre
www.physics.adelaide.edu.au/cssm/workshops/inpc2016/

**18–23 September 2016**
The changing face of galaxies: uncovering transformational physics
Wrest Point Hotel, Hobart
www.caastro.org/event/2016-galaxy

**19–23 September 2016**
ICEAA-2016: International Conference on Electromagnetics in Advanced Applications. Cairns, Queensland
www.iceaa-offshore.org/j3/

**19–23 September 2016**
IEEE-APS Topical Conference on Antennas and Propagation in Wireless Communications (APWC). Cairns, Queensland
www.iceaa-offshore.org/j3/

**16–19 October 2016**
SPIE Bio-Photonics Australasia 2016. Adelaide Convention Centre
spie.org/conferences-and-exhibitions/bio-photonics-australasia

**4-8 December 2016**
22nd AIP Congress - in association with the 13th Asia Pacific Physics Conference. Brisbane Convention Centre
aip-appc2016.org.au